# Deep Learning to Improve the Discovery of Near-Earth Asteroids in the Zwicky Transient Facility

Belén Yu Irureta-Goyena[1], George Helou[2], Jean-Paul Kneib[1], Frank Masci[2], Thomas Prince[3], Kumar Venkataramani[2], Quanzhi Ye (叶泉志)[4,5], Joseph Masiero[2], Frédéric Dux[1], and Mathieu Salzmann[6]
[1] Laboratory of Astrophysics, École Polytechnique Fédérale de Lausanne, Chemin Pegasi 51, 1290 Versoix, Switzerland; belen.irureta@epfl.ch
[2] IPAC, California Institute of Technology, MS 100-22, Pasadena, CA 91125, USA
[3] Division of Physics, Mathematics and Astronomy, California Institute of Technology, Pasadena, CA 91125, USA
[4] Department of Astronomy, University of Maryland, College Park, MD 20742, USA
[5] Center for Space Physics, Boston University, 725 Commonwealth Ave, Boston, MA 02215, USA
[6] Swiss Data Science Center and CVLab, École Polytechnique Fédérale de Lausanne, Rte Cantonale 1015 Lausanne, Switzerland
*Received 2025 January 23; revised 2025 March 27; accepted 2025 April 9; published 2025 May 21*

## Abstract

We present a novel pipeline that uses a convolutional neural network (CNN) to improve the detection capability of near-Earth asteroids (NEAs) in the context of planetary defense. Our work aims to minimize the dependency on human intervention of the current approach adopted by the Zwicky Transient Facility (ZTF). The target NEAs have a high proper motion of up to tens of degrees per day and thus appear as streaks of light in the images. We trained our CNNs to detect these streaks using three datasets: a set with real asteroid streaks, a set with synthetic (i.e., simulated) streaks and a mixed set, and tested the resultant models on real survey images. The results achieved were almost identical across the three models: $0.843 \pm 0.005$ in completeness and $0.820 \pm 0.025$ in precision. The bias on streak measurements reported by the CNNs was $1.84 \pm 0.03$ pixels in streak position, $(0.817 \pm 0.026)°$ in streak angle and $-0.048 \pm 0.003$ in fractional bias in streak length (computed as the absolute length bias over the streak length, with the negative sign indicating an underestimation). We compared the performance of our CNN trained with a mix of synthetic and real streaks to that of the ZTF human scanners by analyzing a set of 317 streaks flagged as valid by the scanners. Our pipeline detected 80% of the streaks found by the scanners and 697 additional streaks that were subsequently verified by the scanners to be valid streaks. These results suggest that our automated pipeline can complement the work of the human scanners at no cost for the precision and find more objects than the current approach. They also prove that the synthetic streaks were realistic enough to be used for augmenting training sets when insufficient real streaks are available or exploring the simulation of streaks with unusual characteristics that have not yet been detected.

*Unified Astronomy Thesaurus concepts:* Near-Earth Objects (1092); Asteroids (72); Surveys (1671)

## 1. Introduction

Near-Earth asteroids (NEAs) can threaten life on Earth. By definition, NEAs are asteroids with an orbit within 1.3 au from the Sun and can thus pass close or even impact our planet. NEAs are divided into four main groups: Amors, which have orbits that do not enter the Earth's orbit; Apollos, which have Earth-crossing orbits that are larger than the Earth's orbit; Atens, which have Earth-crossing orbits that are smaller than the Earth's orbit, and Atiras, which have orbits inside the Earth's orbit.[7] The Chicxulub impactor that led to the extinction of the dinosaurs was estimated to be an NEA ∼10 km in size (Alvarez et al. 1980; Schulte et al. 2010). Other examples of NEA impactors have followed: the largest recent event recorded occurred in 1908, when an impactor exploded near the city of Tunguska, in Siberia. With an estimated diameter of ∼50 m, the Tunguska impactor released enough energy to destroy an area of 2150 km$^2$ of forest (Chyba et al. 1993).

To foresee these events and attempt to mitigate their consequences, we must fully account for all NEAs and their orbital elements. However, despite large NEAs the size of the Chicxulub impactor being well studied, most small-to-medium NEAs have yet to be discovered, and as much as 99% of the estimated NEA population larger than 1 m is still unknown (Harris & D'Abramo 2015). The reason for this is that although small objects dominate the size distribution, they are only detectable when they approach the Earth and appear brighter and with high proper motions.

---

[7] https://cneos.jpl.nasa.gov

More specifically, at high proper motions, which can range up to tens of degrees per day, NEAs can leave signature streaks of up to several hundred pixels, depending on the pixel size and the exposure time. Scouting for these streaks in a range of astronomical images can help us discover new NEAs and refine the orbital elements of known ones.

### 1.1. State of the Art in Asteroid Streak Detection

Detecting these streaks of light, trailed images of the NEAs, was previously done using straightforward image-processing techniques, such as line-detection algorithms. Both the Hough and Radon transform translate the objects of interest into a parameter space where their linear structure is recognizable (Nir et al. 2018). However, these algorithms often mistake the asteroid streaks with other linear objects of comparable size in the image, such as diffraction spikes, and thus require additional detection steps. Similarly, techniques that perform background subtraction to find transient objects can be helpful as intermediate steps toward streak detection but are insufficient for separating asteroid streaks from other elongated features. Specific software has been developed for streak detection, such as the European Space Agency's StreakDet (Virtanen et al. 2016), which relies on a combination of eigenvalue analysis, the Hough transform and a moving 2D Gaussian approach. StreakDet has been applied for asteroid detection in Euclid and VLT Survey Telescope (VST) images (Pöntinen et al. 2020; Saifollahi et al. 2023) but shows a high false-positive rate in single exposures. This implies that for effective detection, a large volume of detections needs to be verified visually or filtered out across multiple exposures of the same field, a requirement often incompatible with the observing strategy of the telescope.

Machine-learning models have surged in the last decade to match and even outperform traditional algorithms for astronomical applications, including streak detection. Some machine-learning algorithms, such as the version presented in Duev et al. (2019), are not capable of detecting the streaks directly and are instead used to confirm the potential candidates previously detected by conventional methods. Compared to standalone traditional algorithms, this approach can speed up the streak-detection process and reduce the need for human inspection. Other algorithms, such as convolutional neural networks (CNNs), are well-suited for directly detecting the streaks since they are designed to extract spatial patterns and features from the images. By undergoing a training process, the CNNs learn to distinguish the streaks left by real objects from other linear features. This approach has been applied to detect asteroid streaks in images taken with the Hubble Space Telescope (Kruk et al. 2022) or in simulated images from the Euclid space telescope (Lieu et al. 2019; Pöntinen et al. 2020), as well as in Irureta-Goyena et al. (2025), which presents the results of applying a CNN trained on synthetic streaks to find asteroids in images taken by the VST.

The present work implements a novel machine-learning pipeline to improve the results of the current official pipeline of the Zwicky Transient Facility (ZTF), mainly described in Duev et al. (2019) and Ye et al. (2019). We should note that in addition to the algorithms developed by the ZTF group, another effort has been made to find asteroid streaks on the ZTF images, discussed in Wang et al. (2022) and focused on faint and short streaks. In the work of Wang et al. (2022), they fed image cutouts to a CNN, which was able to determine whether each cutout contained a streak or not. However, the CNN used could not give any information on the position, angle or length of the streak, which had to be identified by hand and subsequently refined using a Markov-Chain Monte Carlo (MCMC) fitting. Requiring visual inspection of each streak to give the approximate guess needed for the fitting represents a significant barrier to automating the detections and processing many images at a time. For this reason, we will not compare our approach with the results from Wang et al. (2022) and will focus on the comparison with the current ZTF pipeline for NEO discovery, herein NEOZTF pipeline.

### 1.2. The ZTF Streak-detection Process

ZTF (Bellm 2014; Bellm et al. 2019; Graham et al. 2019; Dekany et al. 2020) is a wide-field optical time-domain survey focused on detecting transient sources, scanning the northern sky at a cadence of $\sim 3760 \, \text{deg}^2 \, \text{hr}^{-1}$. It uses the Samuel Oschin Telescope located on Palomar Mountain, California, and is operated by Caltech. Data processing and archiving occur at IPAC, located on the Caltech campus (Masci et al. 2018). The ZTF camera is comprised of 16 CCD chips spanning $\simeq 47 \, \text{deg}^2$, where each pixel has an angular size of $\approx 1''$. The nominal exposure time is 30 s ($\sim 98\%$ of images in the archive).

The NEOZTF effort pioneered systematic detection and reporting of fast-moving asteroids with trailed images while also conducting classic point-source-based detection of slower-moving asteroids in all difference images generated by ZTF (Masci et al. 2018). The streak-extraction process for detecting fast-moving objects is run as part of the nightly production pipeline. During production, raw images are instrumentally calibrated and then pushed to the image-differencing step. This step optimizes transient detection by suppressing both stationary and flux-stable sources in each direct CCD image. The process subtracts from each image a "reference" image of the same field that was constructed by co-adding high-quality images acquired earlier in the survey (for details, see Masci et al. 2018). The resulting difference images contain only the sources that changed in position or flux relative to the input reference image. Additionally, bad or unusable pixels, bright-source artifacts and saturated regions are masked prior to source extraction. Following image differencing, transients such as supernovae, variable stars, gravitationally lensed quasars, comets and asteroids become more easily discernable against the static astrophysical sky, particularly in regions with complex backgrounds and high source confusion. Figure 1 shows an example of the same image containing a streaking asteroid before and after undergoing image differencing. The streak is more easily distinguishable in the latter.

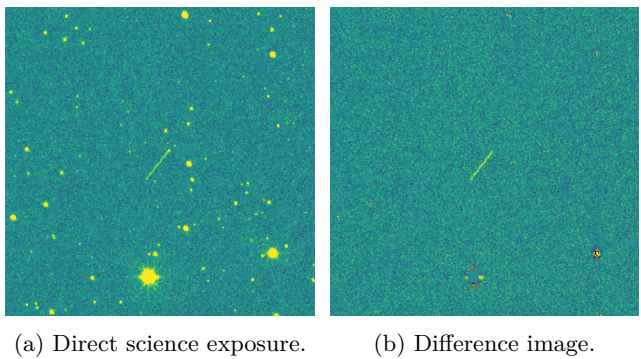

(a) Direct science exposure.　　(b) Difference image.

**Figure 1.** After applying difference imaging, static sources are suppressed from the field. Transient objects, such as streaking asteroids (center of the image) become more discernable against the background.

Difference images are then fed into the first step of the streak-extraction process: the extraction of pixel signals with an approximately linear morphology. These are referred to as raw streaks or the raw candidates for fast-moving objects. Currently, since all linear objects are picked up by the algorithm, over 99.9% of the raw detection stream are identified with false positives (Duev et al. 2019). These are mostly due to imperfect subtractions arising from a combination of variations in photometric throughput (flat-fielding) and astrometry (optical distortion) on short timescales. They can also be due to unmasked cosmic rays, optical artifacts (e.g., diffraction spikes), or long trails from artificial satellites and aircraft.

The raw streaks are then fed to a "stretched-PSF"-fitting routine, which fits each streak with a line connecting the notional endpoint positions and width equal to the full width at half maximum (FWHM) of the image seeing. This fitting results in estimates of the integrated flux, refined endpoint positions, length and deviation from linearity of the morphology of the input streak. These metrics are then thresholded to mitigate the most obvious spurious extractions from the raw stream. This filtering results in a reduction of ∼10% in the number of false positives.

The filtered raw streaks are then staged and pulled by the ZTF-Streak marshaling system (ZSTREAK). This pulling occurs continuously as data are processed throughout the night. The ZSTREAK system feeds each incoming image cutout of a streak into the *DeepStreaks* machine-learned classifier. This classifier is based on a combination of image-based classifiers integrated into a deep-learning framework, detailed in Duev et al. (2019). This classifier further reduces the number of false positives, typically by at least ∼90%. The number of streaks remaining and ready for human (visual) scanning grows to typically a few thousand by the end of each night. The streaks selected by the human scanners as viable candidates undergo further checks to ensure they are not associated with artificial satellites. The remaining streaks are then compared to other human-vetted streaks from earlier in the night or a recent night to check if they can be associated with the same object. If a credible sun-centered orbit is found to fit the streaks simultaneously, the associated streak endpoint positions and observation times are submitted to the MPC. The end-to-end process yields approximately ten asteroids per night in total, which includes already known (cataloged) objects. Following six years of operations, ZTF has discovered over 300 NEOs using the streak-extraction/ZSTREAK system alone.

The system described above can be optimized and made more efficient in two areas: first, during the initial streak-detection process (which yields an extremely high number of false positives); second, there is a need for human scanners to visually examine hundreds to a few thousand streaks every night. The motivation for the work presented here is to explore an automated streak-processing and vetting system that could complement or extend the current operational system. More specifically, our goal is to develop a method that can: (i) decrease the number of spurious extractions during early processing, (ii) reduce the dependence on realtime human scanning and (iii) continue to discover or recover a comparable number of asteroids and possibly more as the survey continues. This framework could also be adapted to future planned surveys.

## 2. Methodology

For the main detection stage of our pipeline, we used a convolutional neural network tailored explicitly for object segmentation, TernausNet (Iglovikov & Shvets 2018), in the implementation developed by Pantoja-Rosero et al. (2022).[8] This architecture, which incorporates a pre-trained block, has been shown to excel when detecting linear image features. The same network was applied in Irureta-Goyena et al. (2025) for direct science exposures of the VST instead of difference images. A schematic of TernausNet is depicted in Figure 2: the neural network is fed an image and produces a detection heatmap of the same size. This is achieved by first reducing the spatial dimensions and increasing the number of channels of the input image, which contain the patterns conveying its features. After each step of this first stage, the network can extract more intricate patterns. In the second stage, the resultant representation of the image is progressively upscaled back to its original size. The network produces a segmentation map through the process where each pixel is assigned a value proportional to its predicted distance from an object of interest; in this case, a streak. Since TernausNet is designed to process PNG images, the FITS files were compressed using a dynamic scale, Astropy ZScaleInterval. After testing the different parameters, the contrast of the interval was set to 0.5 to enhance the faint objects while not oversaturating the brighter ones. While the conversion from FITS to PNG images entails compression, if done adequately, reducing the image range could filter down the amount of information provided to the CNN and help it focus on the important features; in this case, the fainter asteroids, which are more difficult to detect.

---
[8] Available at https://github.com/bgpantojar/topo_crack_detection.

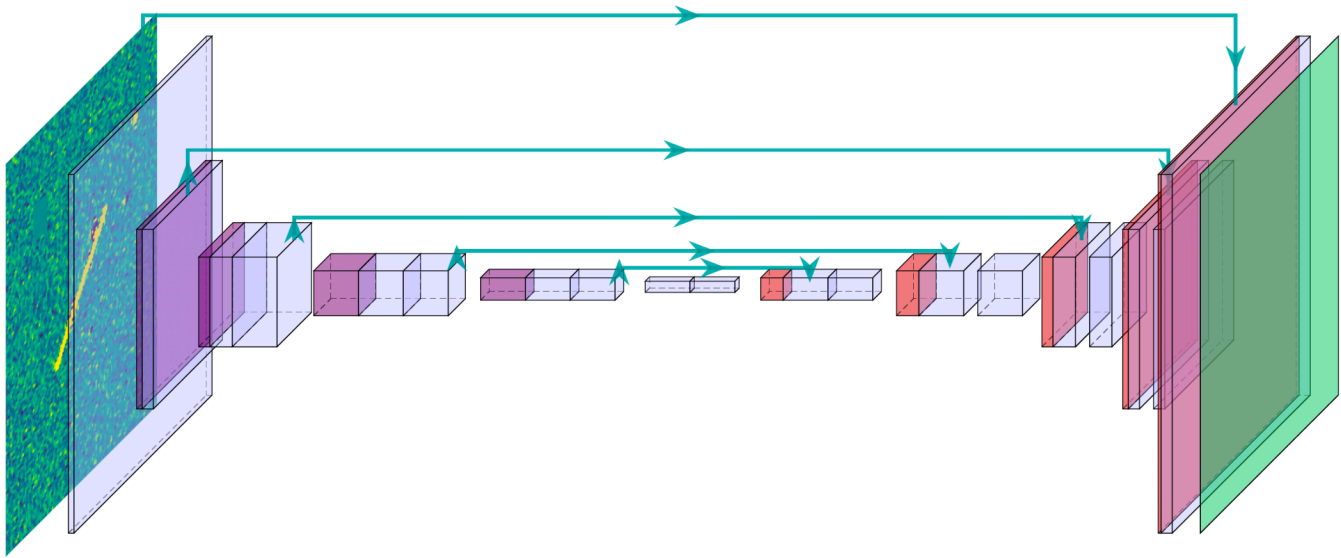

**Figure 2.** TernausNet architecture, adapted from Irureta-Goyena et al. (2025), Iglovikov & Shvets (2018). Each block represents a transformation applied to the image: white for convolutions and transpose convolutions, purple for max pooling and red for unpooling. The thickness and height of the blocks are proportional to the resultant number of channels and map size, respectively. Green arrows show skip connections. Adapted from Irureta-Goyena et al. (2025) & Iglovikov & Shvets (2018). CC BY 4.0.

As a probability map, the detection heatmap yields no explicit knowledge of how many asteroids have been detected in an image. For that, the heatmap is converted to a binary map by setting a threshold probability above which a pixel is considered part of a detection. In practice, the choice of the threshold will influence the number of detections. A lower threshold will yield more streaks but is more likely to produce false positives. On the contrary, a higher threshold will risk rejecting real detections but will not mistakenly accept linear features that are not real streaks. Figure 3 shows the same streak converted to a binary map using two different thresholds. A detailed discussion on the choice of threshold is found in Section 3.1.

### 2.1. Endpoint Fitting

The exact position of the streaks found is essential to perform astrometric analysis and is required for submitting the detections to the Minor Planet Center (MPC). Once the detection heatmaps were converted to binary maps, the contours in each image were fitted using a preliminary linear fit on the binary map, which yielded an initial position of the two endpoints of the streak. These endpoints were used as an initial guess for a more exhaustive fit on the original FITS, using an MCMC algorithm in the implementation described in Hogg & Foreman-Mackey (2018), with 5000 random walks per fit. A higher number was suggested not to improve the fitting results while being computationally expensive.

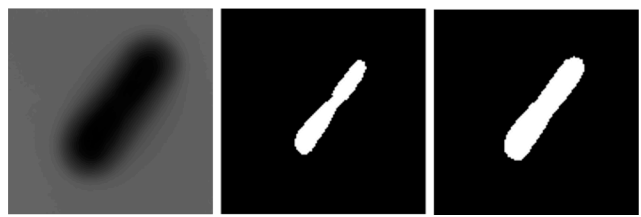

**Figure 3.** Representation of a streak at different thresholds. From left to right: streak in the detection heatmap, binary map at the threshold of pixel value 13, binary map at the threshold of pixel value 20.

### 2.2. Training Data

The images used for this study were real ZTF difference images taken from 2018 to 2023. They were of size $3072 \times 3080$ pixels, equivalent to an angular area of $\approx 0.75 \, \text{deg}^2$, and had been created by the NEOZTF pipeline by dividing each CCD of the original image into four quadrants. For the training, three datasets were used: a set with real asteroids, a set with synthetic asteroids, injected into the real images, and a set containing a mix of synthetic and real asteroids. The set with real asteroids comprised 450 cutouts of asteroids detected by the pipeline currently in operation and confirmed by the MPC. The images selected were in the three custom ZTF filters: the g-filter (≈31%), the i-filter (≈2%) and the r-filter (≈67%), a discussion of which can be found in Masci et al. (2018). Their seeing distribution followed the normal distribution of the ZTF frames, centered around ≈2″. The distribution of the asteroid streaks in terms of apparent magnitude and streak length can be found in Figure 4. The upper limit in streak length of 160 pixels corresponds to a proper motion of $5\rlap{.}{''}33\,\mathrm{s}^{-1}$. The values were taken from the NEOZTF catalog, which in a minority of cases could contain inaccuracies, such as the one-pixel streaks, which, upon visual inspection, appeared several pixels long. However, inaccuracies were deemed exceptional and thus did not significantly affect the training process.

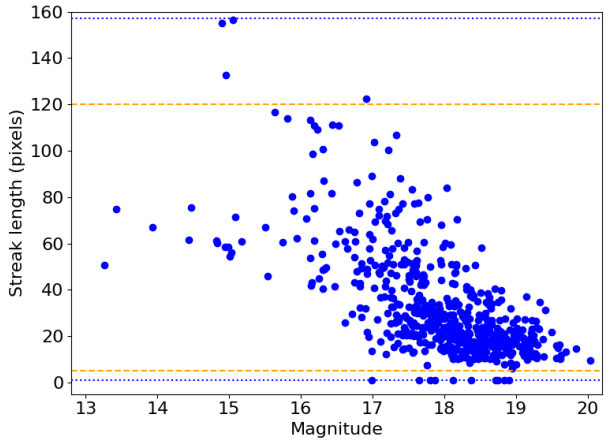

**Figure 4.** Distribution of the real asteroids used in the training and testing datasets, in terms of magnitude and length of streak. These values were determined by the NEOZTF pipeline. The blue dotted line depicts the range of the real distribution (1–157 pixels) and the orange dashed line, the range of the synthetic distribution (5–120 pixels).

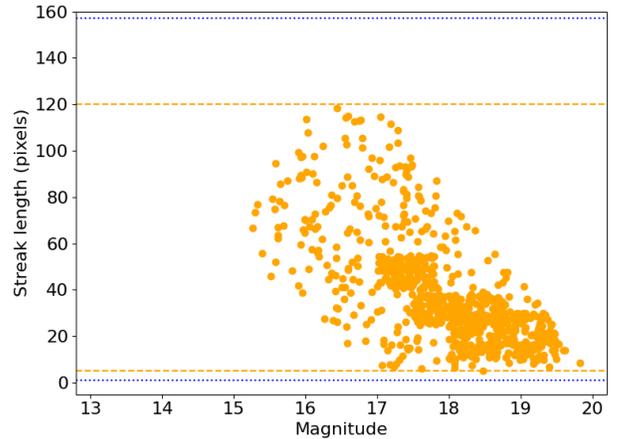

**Figure 5.** Distribution of the synthetic asteroids used in the training and testing datasets, in terms of magnitude and length of streak. For clarity, only a fraction of the ≃20,000 asteroids is depicted. These values were established before injecting the asteroids into the frames. The blue dotted line depicts the range of the real distribution (1–157 pixels) and the orange dashed line, the range of the synthetic distribution (5–120 pixels).

Although past detections can provide an accurate representation of the target asteroids, if used for training the CNN, they could skew the new pipeline and restrict it to asteroids identical to those already found. A new, more comprehensive set of asteroids was created with the aim of finding streaks that could have been missed by the current pipeline. This simultaneously addressed the lack of training examples and the potential biases in the model.

A population of ≃20,000 synthetic asteroids was injected into ZTF difference images that did not contain asteroids, chosen randomly and without any other differences from the images with real asteroids. The magnitude and streak length were crafted to approximately follow that of the real asteroids, but, as displayed in Figure 5, attention was put to filling in some gaps in the current distribution to test whether the new pipeline could reach beyond the coverage of the ZTF approach. No synthetic asteroids were placed at the brightest end of the real asteroid distribution since the very bright asteroids are generally well-known and easier to detect. Instead, the focus was placed on the new algorithm to detect the fainter end of the distribution. The streaks had lengths 5–120 pixels, with the lower limit set to avoid confusion with static but extended sources, such as galaxies, and the upper limit set by following the real upper limit but disregarding the outliers. Given the exposure time of 30 s, the proper motions depicted ranged from $0\rlap{.}{''}17\,\mathrm{s}^{-1}$ to $4\rlap{.}{''}00\,\mathrm{s}^{-1}$.

To make the appearance of the streaks realistic, each injected streak was convolved with the individual PSF of each frame. Figure 6 shows the comparison between synthetic and real streaks of comparable magnitudes and streak lengths in images of similar seeing. The synthetic streaks appear virtually indistinguishable from the real streaks.

Lastly, the real and the synthetic asteroids were combined in a mixed dataset. This aimed to build a dataset trained with the most realistic examples but able to overcome current biases. All training sets contained examples of those features that usually prompt false positives in NEOZTF pipeline, as shown in Figure 7. They were not labeled as positive examples, so the model would learn not to flag them as asteroids mistakenly.

### 2.3. Model Development and Assessment

To isolate the effect of the training dataset on the model performance, three models, one per dataset, were trained in

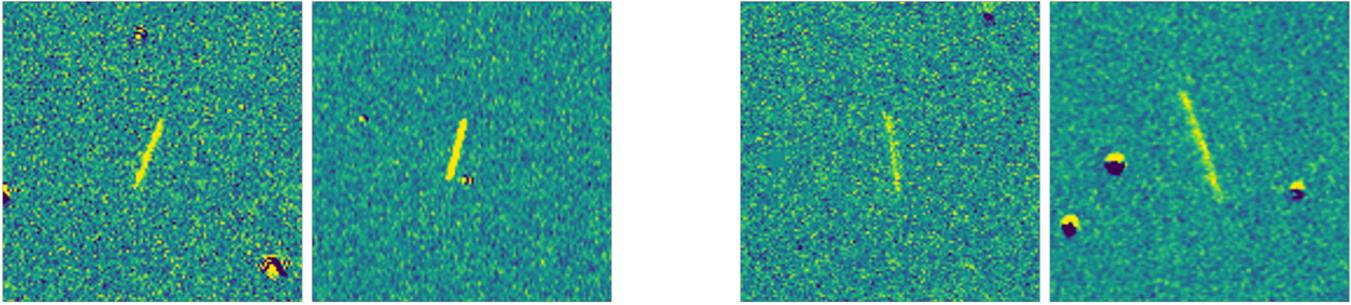

(a) Left: synthetic asteroid, magnitude 17.92, length 43 pixels, seeing 1.74″. Right: real asteroid, magnitude 17.88, length 29 pixels, seeing 1.74″.

(b) Left: synthetic asteroid, magnitude 18.00, length 37 pixels, seeing 4.78″. Right: real asteroid, magnitude 17.80, length 50 pixels, seeing 4.44″.

**Figure 6.** Example comparisons between synthetic streaks and real asteroids of similar characteristics. The magnitude and length of the real streaks are provided as reported by the ZTF team.

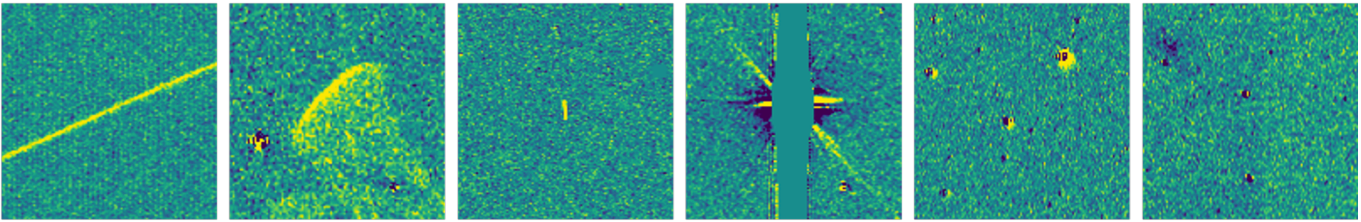

**Figure 7.** Examples of linear features that can trigger false positives in the current NEOZTF pipeline. These were included in the training set of our pipeline and not labeled as asteroids so the algorithm would learn not to confuse them. From left to right: satellite, "dementor," cosmic ray, masked bright star, incorrect subtraction and "ghost." Adapted with permission from Duev et al. (2019) © 2019 The Author(s) Published by Oxford University Press on behalf of the Royal Astronomical Society.

identical conditions: during 32 epochs, using a batch size of 16 and a learning rate of $5 \times 10^{-6}$. This combination of hyperparameters was suggested in Pantoja-Rosero et al. (2022) to minimize the loss for this network. Once trained, the models were applied to the same test set, which comprised 115 images containing real NEAs that had not been used for training and, thus, had not been seen by the algorithm before. Since the pipeline was devised to find real streaks, the performance of the models when finding synthetic streaks was irrelevant to this assessment. The asteroids in the test set had been detected by the current NEOZTF pipeline and confirmed as real asteroids by the MPC. The ground-truth position, orientation and length of the streaks were provided by the ZTF team and used to analyze the results. A detection with the new pipeline was considered valid if the initial guess for the midpoint position of the streak was strictly less than 5 pixels— approximately twice the seeing—away from the ground truth. As discussed in Section 2.1, this coarse value would be subsequently refined.

The performance was compared by assessing three indicators, well-established in the state of the art:

$$\text{Completeness} = \frac{\text{TP}}{\text{TP + FN}}, \quad (1)$$

$$\text{Precision} = \frac{\text{TP}}{\text{TP + FP}}, \quad (2)$$

$$\text{F1} = \frac{2 \times \text{Completeness} \times \text{Precision}}{\text{Completeness + Precision}}, \quad (3)$$

where TP corresponds to true positives, FP to false positives and FN to false negatives. As discussed in Irureta-Goyena et al. (2025), the F1 score, which represents the balance between completeness and precision, was evaluated at each threshold to determine the optimal thresholding value.

Beyond testing the performance across models, the new pipeline output was compared to the current NEOZTF output after it has gone through detection review and confirmation by human scanners. A set of 303 images taken in 2024 June and containing detections flagged by the scanners were fed to the new pipeline. This data set will be called the scanners data set hereafter. Two aspects were analyzed:

1. The completeness of the new pipeline was evaluated in relative terms, computing the fraction of detections by human scanners also found by the algorithm.
2. Any additional detections by the new pipeline that had not been flagged by the scanners were sent to the member of the ZTF team who verifies the detections from the scanners. This member confirmed whether the additional detections were false positives or correct detections.

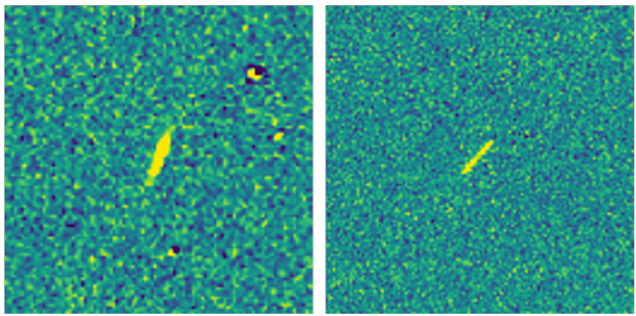

**Figure 8.** Left: confirmed streak from a fast-rotating artificial object. Right: confirmed real asteroid.

At this stage of the NEOZTF pipeline, the streaks left by asteroids and the streaks left by fast-rotating artificial objects, namely satellites and space debris, are considered indistinguishable since they can have the same visual appearance, as displayed in Figure 8. Human scanners are asked whether the streak belongs to a real object, and it is not until the astrometry of the object has been resolved at a later stage that the asteroids are recognized as credible candidates. Since we compared our pipeline with the NEOZTF pipeline up until that phase, we use the same convention when describing our results.

## 3. Results and Discussion

### 3.1. Results of Testing Across Models

This subsection presents the details and results of the model development. The F1 scores after applying the three models, trained with synthetic, real and mixed streaks, were evaluated at different thresholds for the binary conversion to determine the optimal threshold for each case, as shown in Figure 9. For both the model trained with synthetic and the model trained with mixed streaks, the curve was almost horizontal in the studied range and peaked at low pixel values, 8 and 10, respectively. The curve for the model trained with real streaks had a steeper slope up to a threshold of ∼20, then flattened out or climbed slowly for the whole range of thresholds tested. This indicates that the model trained with real streaks showed a higher gradient of uncertainty in the detection heatmap than the other two, possibly due to the exposure to fewer example streaks during the training, particularly in the shorter end of the distribution (∼10 pixels).

The results when using each optimized value for every model are shown in Table 1. They are almost identical across the three models, with the mixed model showing slightly better completeness and slightly worse precision than the other two. This is explained by the fact that, generally speaking, the mixed model was merging all hits from the other two models,

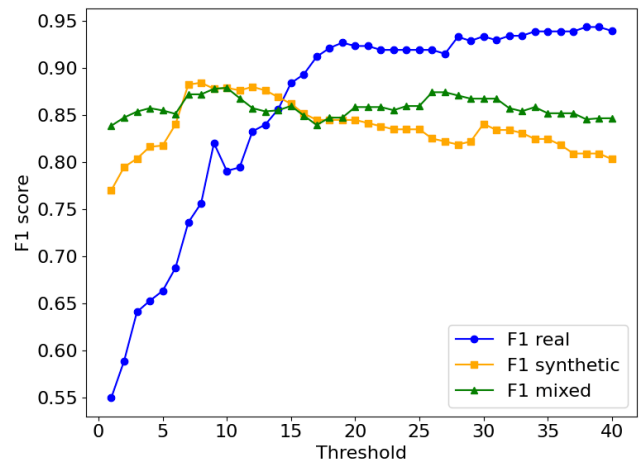

**Figure 9.** F1 scores at different thresholds (pixel values) of the models trained with the real, the synthetic and the mixed datasets. The F1 score is evaluated before the endpoint fitting.

**Table 1**
Comparison of the Performance Indicators Defined in Section 2.3 For the Tests with the Three Different Models

| Metric | Real Set | Synthetic Set | Mixed Set |
|---|---|---|---|
| Total hits | 114 | 118 | 124 |
| True positives | 97 | 96 | 98 |
| False positives | 17 | 22 | 27 |
| False negatives | 18 | 19 | 17 |
| **Completeness** | **0.84** | **0.84** | **0.85** |
| **Precision** | **0.85** | **0.82** | **0.79** |
| F1 score | 0.85 | 0.83 | 0.81 |

**Note.** See Equations 1–3 for definitions. The F1 score is evaluated after the endpoint fitting.

both the correct detections and the false positives. Since the mixed model was trained with all the example streaks of the two other models combined, this was expected. The standard deviation across the values is 0.005 for the completeness and 0.025 for the precision.

The asteroid streaks not found by the detection models are displayed in Figure 10 in order of ascending magnitude and within boxes of 128 × 128 pixels. The brightest streaks were missed or split in two in most cases, possibly due to a negative bias against them in the three training sets (see Figures 4 and 5). In the particular case of the synthetic model, as discussed in Section 2, no streaks brighter than magnitude 15.25 had been modeled. Fainter streaks were, in most cases, either shown in the heatmap but neglected after the thresholding step or inaccurately positioned during the endpoint fitting, thus discarded for being more than 5 pixels away from the ground truth. No correlation was found between detection success and the seeing of the frames.

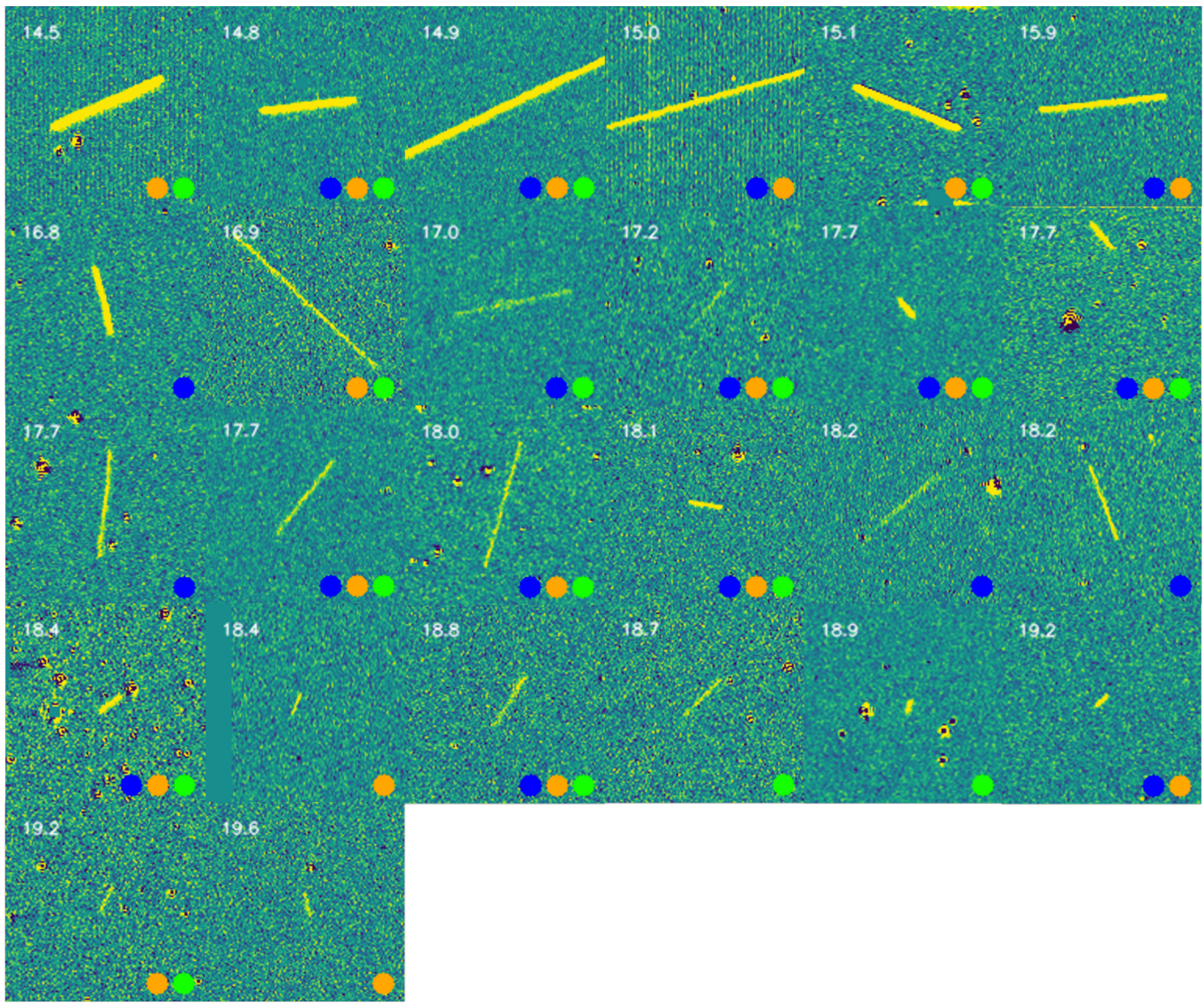

**Figure 10.** Streaks missed by each of the models. The circles in the bottom right corner indicate which of the models did not find that particular streak, with blue, orange, and green, respectively, corresponding to the real, the synthetic and the mixed models. The values in the top left corner indicate the apparent magnitude of the streak, as estimated by the NEOZTF pipeline.

Figures 11–13 show the bias of the pipeline when characterizing the streaks found by each of the three models. For an adequate comparison of the graphs, the points plotted have been averaged over a similar number of values and thus appear unevenly spaced. Figure 11 displays the bias in the position of the streak midpoint, defined as the absolute distance in pixels between the ground truth and the detection, as a function of magnitude and streak length. Beyond one stray point at a lower value in the real model of Figure 11(b), no conclusive correlation can be stated between any of the variables considered and the bias in streak position. Figure 12 displays the bias in the angle with the $x$-axis of the detected streak, computed using the absolute value, as a function of magnitude and streak length. Although it shows no dependency on the type of training data or the magnitude of the streak, the bias in angle is higher for short streaks, since in these cases, any bias in the endpoint determination will disproportionally affect the angle calculated. Lastly, Figure 13 depicts the fractional bias in streak length, calculated as the absolute bias in streak length divided by the streak length, with a positive or a negative sign indicating that the length was overestimated or underestimated, respectively. Figure 13(a) shows that the fainter streaks were often underestimated, an expected trend since they can be blurrier near the edges and shortened in the detections. As a function of the streak length, shown in Figure 13(b), the fractional bias in length was more

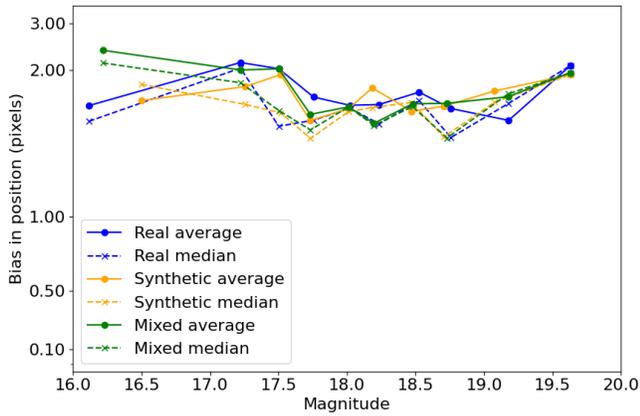

(a) Bias in the detected position vs streak magnitude.

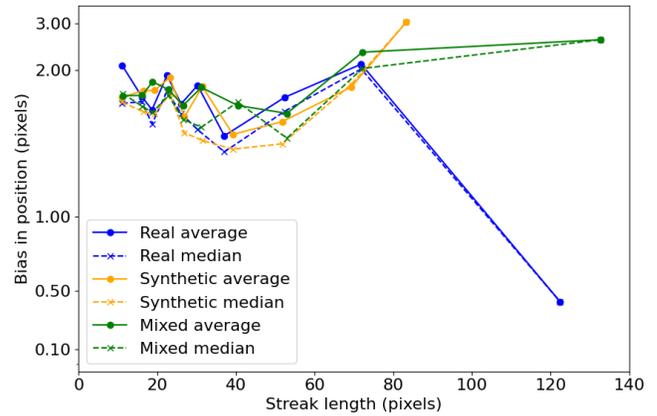

(b) Bias in the detected position vs true streak length.

**Figure 11.** Bias in the detected position of the streak as a function of the type of training set.

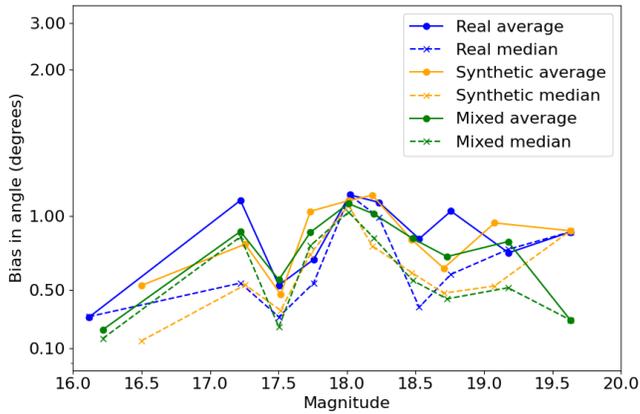

(a) Bias in the detected angle vs streak magnitude.

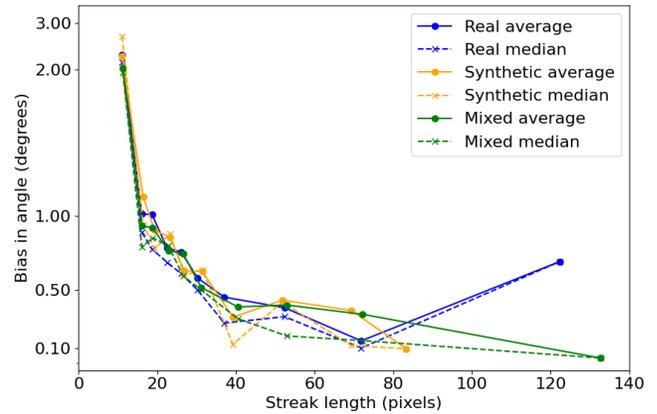

(b) Bias in the detected angle vs true streak length.

**Figure 12.** Bias in the detected angle of the streak as a function of the type of training set.

acute in the case of the shorter streaks, whose length was also underestimated. However, this trend was a direct result of using relative values, since the absolute bias in length was consistent throughout the whole length distribution, but yielded a higher fractional bias when compared to the length of the short streaks. No difference between the three models could be inferred from the streak length determination either.

A statistical accounting of the differences in characterization across models can be found in Table 2. The quantitative analysis confirms that the three models performed almost equally in all respects: the standard deviation across models for the biases in position and angle, and the fractional bias in streak length were 0.03 pixels, 0°.026 and 0.003, respectively. Together with the results discussed in Figures 11–13, these values prove that the model trained with real streaks, the model trained with synthetic streaks and the model trained with the mix were able to find almost the same number of asteroids while keeping approximately the same precision and accuracy when characterizing the detections. From this development, we conclude that the synthetic streaks are to all effects realistic and can thus be used to substitute real asteroids in future training sets, either to enlarge training sets where there are not sufficient real streaks or to explore the simulation of streaks with unusual characteristics (either in terms of magnitude or streak length) that have not yet been detected.

### 3.2. Results on the Comparison with the NEOZTF Pipeline

In the previous subsection, we demonstrated that introducing synthetic asteroids in the training set of real asteroids does not negatively impact the ability of the neural network to detect, at least, the asteroid streaks identifiable by the current

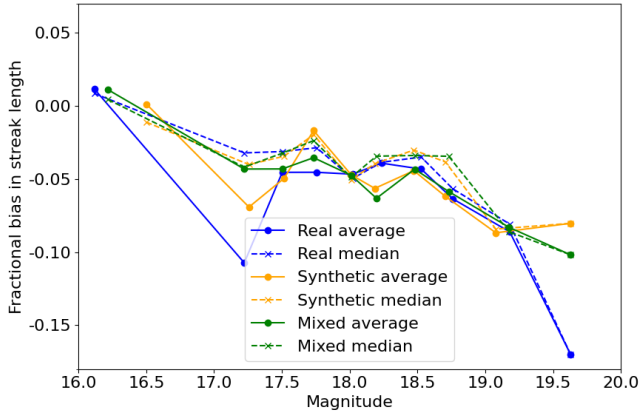
(a) Bias in the detected streak length vs streak magnitude.

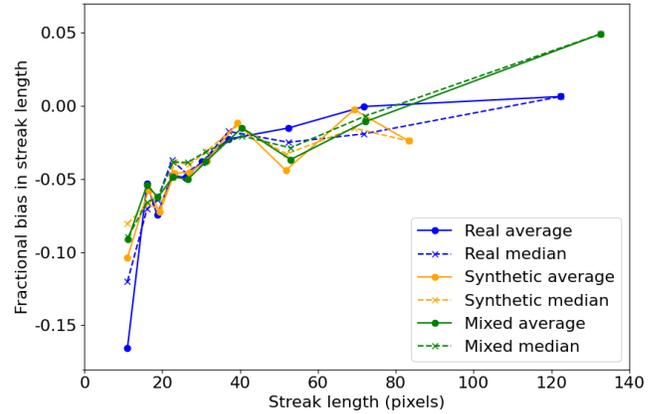
(b) Bias in the detected streak length vs true streak length.

**Figure 13.** Fractional bias in the detected streak length of the streak as a function of the type of training set. The bias was computed by dividing the absolute length bias by the streak length in each case. A positive bias indicates that the length was overestimated, while a negative bias implies that the length was underestimated.

**Table 2**
Accuracy Characterizing the Detected Streaks After Applying Each of the Three Models

| Metric (Average) | Real | Synthetic | Mixed |
|---|---|---|---|
| Bias in position (pixels) | 1.84 | 1.80 | 1.87 |
| Fractional bias in streak length | −0.052 | −0.047 | −0.045 |
| Bias in angle (degrees) | 0.83 | 0.84 | 0.78 |

NEOZTF pipeline. The results in this subsection address (1) whether our models can find as many objects as the human scanners who perform a nightly search on ZTF data and (2) whether introducing synthetic asteroids can help our models find new objects that are being missed by the pipeline in place.

The pipeline using the model trained only with real streaks found 888 streaks in the scanners dataset. This included 70 % of the objects that the scanners had detected, using the same convention as in the previous subsection, where a streak was considered detected if its position was estimated with an error of $<5$ pixels (5″, approximately twice the seeing). The additional detections were submitted to the ZTF member verifying the detections of the scanners, who determined that 10 of the 888 streaks found were false positives and that the rest were valid detections, which corresponds to a precision of $>0.98$.

The pipeline using the model trained with a mix of real and synthetic streaks found a higher number of objects when applied to the scanners dataset, 963 streaks. The performance when detecting the objects that the scanners had detected was also better, since it found 80% of those streaks. The ZTF member verifying the detections reported that 14 out of the 963 streaks were false positives, a precision of $>0.98$. Figure 14 shows several examples of additional detections by the real and the mixed model that were deemed to be real objects. A quantitative accounting is depicted in Table 3.

This disparity in completeness between the mixed and the real models, which was not seen in the test set of 115 real asteroids, could be due to two possible reasons. First, upon individual inspection of the streaks, both the streaks found by the scanners and the additional streaks found by the new pipelines, the majority ($>80\%$) appeared to be artificial satellites, whose detectability plausibly differs from that of natural satellites. This is believed to be the main driver of the disparity. Second, since the 115 real asteroids were identical in distribution to those used for training the real model, it was expected that it would perform adequately in this test set, but its performance could worsen when finding objects of different characteristics. The mixed model, trained with a more diverse dataset, would be better equipped for finding new detections. The precision for both models, higher than that achieved in the test set, could be driven by several factors. The density of real objects per image was three times higher in the scanners set than in the test set, which, for a constant number of false positives per image, would result in a higher precision. In addition, the criterion used for the additional streaks found in the scanners set was more lenient than for the test set since a streak detected was considered valid if deemed credible by the ZTF scanner, regardless of it being an asteroid or a satellite, or not confirmed by the MPC.

The new detections were analyzed in terms of their streak length. As shown in the normalized histograms of Figure 15, no significant differences were found between the distribution of the streak lengths detected by the scanners and those detected by the pipelines with the real and the mixed model. Quantitative magnitude evaluation, which was not included in our pipeline, requires a nontrivial photometric analysis that, given the extended nature of the sources, is highly complex

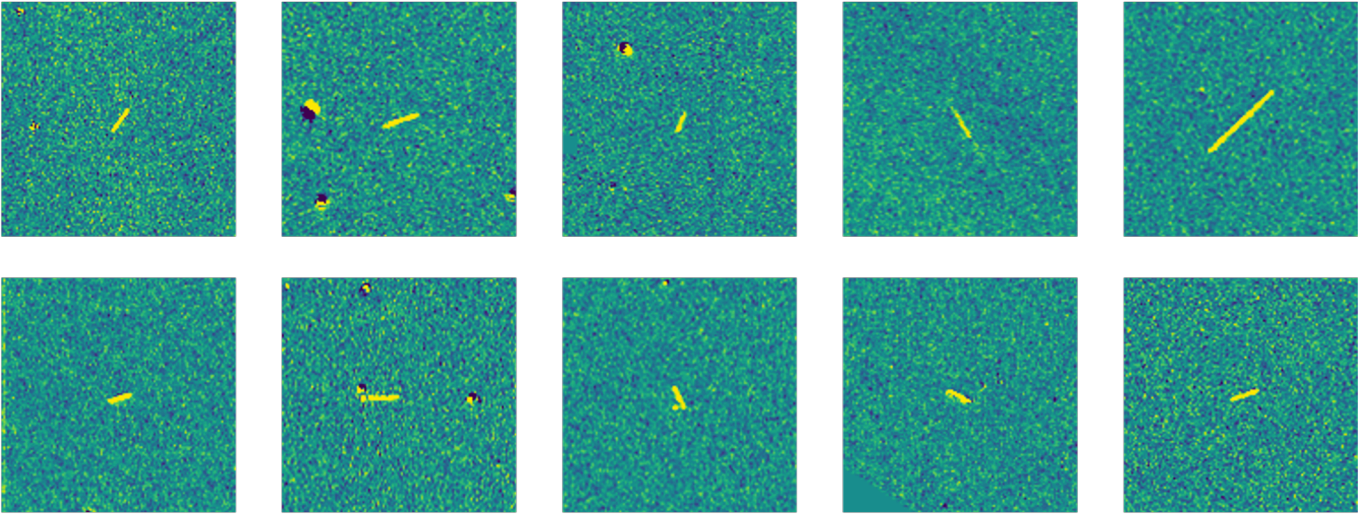

**Figure 14.** Verified detections by the pipeline not originally spotted by the human scanners. The top row shows streaks that were detected by both the real model and the mixed model, while the bottom row displays detections by the mixed model only.

**Table 3**
Performance of the Pipeline by Training Set When Applied to the Scanners Set

| Metric | Real Set | Mixed Set |
| --- | --- | --- |
| Total hits | 888 | 963 |
| Detections also reported by scanners | 222 | 252 |
| True positives | 878 | 949 |
| False positives | 10 | 14 |
| **Completeness of scanners set** | **0.70** | **0.80** |
| **Precision** | **0.98** | **0.98** |
| F1 score | 0.82 | 0.89 |

**Note.** See Equations 1–3 for definitions.

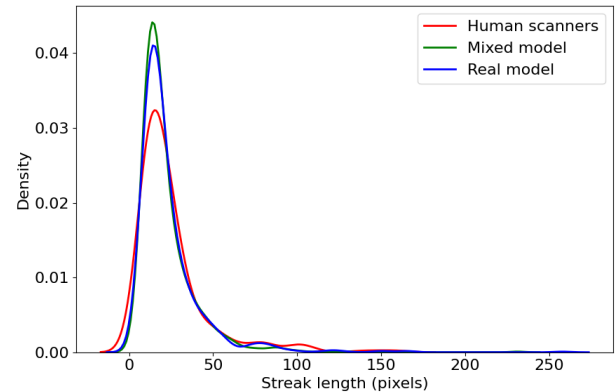

**Figure 15.** Normalized distribution of the streak length in the detections found by the human scanners, the pipeline with the model trained with mixed streaks and the pipeline with the model trained with real streaks.

and goes beyond the scope of this study. The brightness of the detections in each set was visually compared. There were no identifiable trends that could distinguish the detections of the scanners from the detections of the new pipelines.

Considering only the objects marked by the scanners, the completeness achieved by the new pipelines was slightly lower than that achieved in the test set. However, the overall completeness, treated as the number of detections over the number of objects of interest, was considerably higher than that achieved by the scanners. Specifically, the real model was able to find 70% of the 317 objects detected by the scanners and 656 additional streaks that corresponded to either asteroids or fast-rotating artificial objects. The mixed model was able to find 80% of the 317 objects and 697 additional valid streaks. This implies that the new method developed performs a more exhaustive search than the ZTF human scanners with no cost for the precision since the false positives remained at <1%. The main driver of the difference in completeness between the scanners and our CNN was likely to lie at an earlier stage of the NEOZTF pipeline, in the deep neural network used to automatically discard the false positives. As described by Duev et al. (2019), streaks are required to reach a threshold score in a set of three different classifiers to be considered as candidates. A strict thresholding criterion could have filtered out the candidates upstream before they reached the human scanners.

### 4. Conclusions and Future Work

We have presented the development, test and application of a machine-learning pipeline for streak detection in images taken with the ZTF survey. For this, we have compared the performances of three different models: one trained with real asteroid streaks previously detected by the pipeline in operation, one trained with synthetic streaks only and one trained with a mix of the two. When tested on a set of past asteroid detections, the results of the three models were almost indistinguishable from each other regarding detection completeness and precision. The completeness-precision rates for the real, the synthetic and the mixed model were 0.84–0.85, 0.84–0.82 and 0.85–0.79, respectively. No significant differences were found when assessing the quality of the detections; the three pipelines performed equally well when characterizing the streaks found in terms of position, angle or length. The average bias was $1.837 \pm 0.03$ pixels for the streak position, $-0.048 \pm 0.003$ for the fractional bias in streak length and $(0.817 \pm 0.026)°$ for the streak angle with respect to the $x$-axis.

We compared the performance of our automated pipeline to that of the human scanners in the current NEOZTF pipeline by processing a set of images with 317 detections flagged by the scanners. The pipeline trained with a mix of real and synthetic asteroids was able to find 80% of the scanner detections and 697 additional detections. These results suggest that our automated pipeline can complement or even substitute the nightly work of the human scanners, at no cost for the precision of the method and with the advantage of finding more objects than the current approach. They also demonstrate that the synthetic streaks simulated are to all effects realistic and can be used to expand the training set. Our pipeline not only shows a strong potential to make new findings in the ZTF data but also can be scaled up to other wide-field telescopes, setting the stage for fast and automated NEA discoveries in the next generation of astronomical surveys.

Future work will focus on enhancing the pipeline with photometry capability, on analyzing the new detections to verify whether they belong to asteroids in the MPC database and on deploying and applying the new pipeline to a more extensive set of images every night. This will further test the effects of introducing synthetic streaks and help identify trends in the characteristics of the new detections. We aim, with this work, to improve our ability to foresee any future NEA impactors.

## Acknowledgments

Based on observations obtained with the Samuel Oschin Telescope 48-inch and the 60-inch Telescope at the Palomar Observatory as part of the Zwicy Transient Facility project. ZTF is supported by the National Science Foundation under grants No. AST-2034437 and a collaboration including current partners Caltech, IPAC, the Oskar Klein Center at Stockholm University, the University of Maryland, University of California, Berkeley, the University of Wisconsin at Milwaukee, University of Warwick, Ruhr University Bochum, Cornell University, Northwestern University and Drexel University. Operations are conducted by COO, IPAC, and UW. We would like to thank the team at IPAC for supporting this work, in particular Avery Wold, Ben Rusholme, Tracy Chen, Daniel Piña and Alexander Hui. We would also like to thank the anonymous reviewer, Pablo Ramìrez and Francisco Ocaña for their helpful comments. This project has received funding from the European Union's Horizon 2020 and Horizon Europe research and innovation programmes under the Marie Sklodowska-Curie grant agreement Nos 945363 and 101105725, and funding from the Swiss National Science Foundation and the Swiss Innovation Agency (Innosuisse) via the BRIDGE Discovery grant 40B2-0 194729/2.

*Facility:* PO:1.2m.

*Software:* Astropy (Astropy Collaboration et al. 2013, 2018), Emcee (Foreman-Mackey et al. 2013), Numpy (Harris et al. 2020), OpenCV (Bradski 2000), PlotNeuralNet,[9] PyTorch (Paszke et al. 2019), SEP (Barbary 2016), Scipy (Virtanen et al. 2020), SourceExtractor (Bertin & Arnouts 1996).

## ORCID iDs

Belén Yu Irureta-Goyena https://orcid.org/0009-0004-5327-8767
George Helou https://orcid.org/0000-0003-3367-3415
Jean-Paul Kneib https://orcid.org/0000-0002-4616-4989
Frank Masci https://orcid.org/0000-0002-8532-9395
Thomas Prince https://orcid.org/0000-0002-8850-3627
Kumar Venkataramani https://orcid.org/0000-0003-3321-1472
Quanzhi Ye (叶泉志) https://orcid.org/0000-0002-4838-7676
Joseph Masiero https://orcid.org/0000-0003-2638-720X
Frédéric Dux https://orcid.org/0000-0003-3358-4834
Mathieu Salzmann https://orcid.org/0000-0002-8347-8637

---

[9] https://github.com/HarisIqbal88/PlotNeuralNet